\documentclass[aps,prl,reprint,superscriptaddress]{revtex4-2}
\usepackage{graphicx}
\usepackage{bm,amssymb}
\usepackage{amsmath}

\usepackage{pdfpages} 
\usepackage{pgffor} 

\makeatletter
\AtBeginDocument{\let\LS@rot\@undefined}
\makeatother

\def\supplementfilename{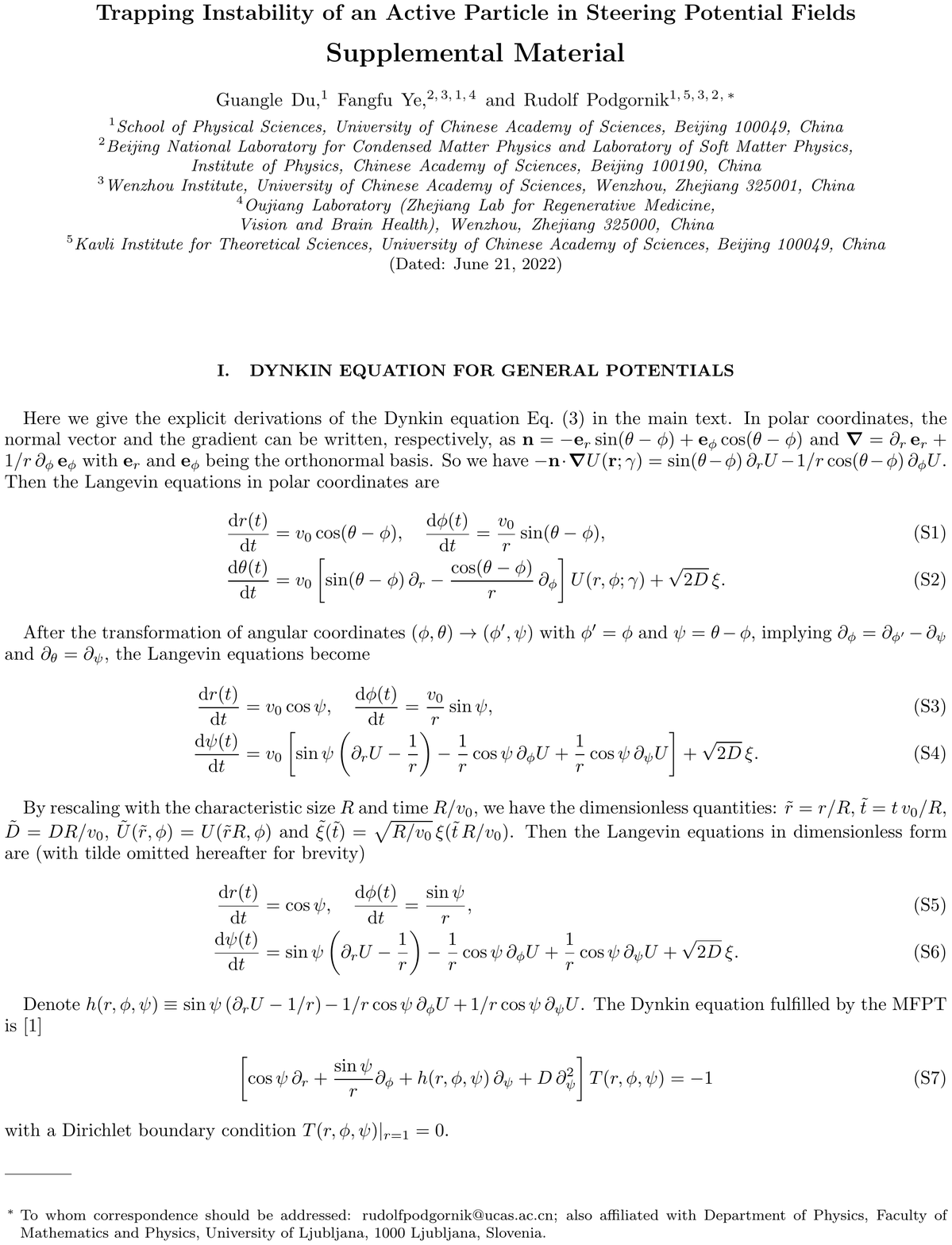}

\pdfximage{\supplementfilename}
\def\numbersupplementpages{\the\pdflastximagepages}

\newif\ifarXiv
\arXivtrue 

\begin{document}
\title{Trapping Instability of an Active Particle in Steering Potential Fields}
\date{\today}

\author{Guangle Du}
\affiliation{School of Physical Sciences, University of Chinese Academy of Sciences, Beijing 100049, China}

\author{Fangfu Ye}
\affiliation{Beijing National Laboratory for Condensed Matter Physics and Laboratory of Soft Matter Physics, Institute of Physics, Chinese Academy of Sciences, Beijing 100190, China}
\affiliation{Wenzhou Institute, University of Chinese Academy of Sciences, Wenzhou, Zhejiang 325001, China}
\affiliation{School of Physical Sciences, University of Chinese Academy of Sciences, Beijing 100049, China}
\affiliation{Oujiang Laboratory (Zhejiang Lab for Regenerative Medicine, Vision and Brain Health), Wenzhou, Zhejiang 325000, China}

\author{Rudolf Podgornik}
\email[To whom correspondence should be addressed: ]{rudolfpodgornik@ucas.ac.cn}
\altaffiliation[also affiliated with ]{Department of Physics, Faculty of Mathematics and Physics, University of Ljubljana, 1000 Ljubljana, Slovenia.}
\affiliation{School of Physical Sciences, University of Chinese Academy of Sciences, Beijing 100049, China}
\affiliation{Kavli Institute for Theoretical Sciences, University of Chinese Academy of Sciences, Beijing 100049, China}
\affiliation{Wenzhou Institute, University of Chinese Academy of Sciences, Wenzhou, Zhejiang 325001, China}
\affiliation{Beijing National Laboratory for Condensed Matter Physics and Laboratory of Soft Matter Physics, Institute of Physics, Chinese Academy of Sciences, Beijing 100190, China}

\begin{abstract}
A particle driven by active self-propulsion can be subject to inhomogeneous potential fields, steering its orientation and leading to confinement and eventual trapping.
Analytical treatment of capture and/or release dynamics for general steering potentials presents a challenge due to its coupling between external potential fields and intrinsic active noise.
By using the projection operator method we obtain the coarse-grained Dynkin equations with orientation integrated out in the large fluctuations limit, and derive explicit analytical solutions for the mean first passage time in radially symmetric point source trapping potentials.
We analyze the ensuing trapping instabilities related to a critical value of the steering potential strength below which the particle either cannot be lured into the trap, or above which it is unable to leave the trap after being lured into it.
\end{abstract}
\maketitle
\paragraph{Introduction.}
\label{sec:org805a823}
Since his seminal work \cite{kramersBrownianMotionField1940}, the {\sl Kramers problem} concerning the confinement and escape of a particle in potential fields with fluctuations has been generalized and elaborated for systems near equilibrium \cite{naehDirectApproachExit1990,hanggiReactionrateTheoryFifty1990,melnikovKramersProblemFifty1991,barcilonSingularPerturbationAnalysis1996}.
Its theoretical extension to self-propelled particles is being mainly treated numerically or by simulations \cite{buradaEscapeRateActive2012,scacchiMeanFirstPassage2018,olsenEscapeProblemActive2020},
but presents an analytical challenge due to the non-equilibrium nature of active systems \cite{pototskyActiveBrownianParticles2012,capriniActiveEscapeDynamics2019,ledoussalVelocityDiffusionConstant2020,ledoussalVelocityDiffusionConstant2020,woillezActiveTrapModel2020,capriniCorrelatedEscapeActive2021}.
While the analytical efforts up till now have been focused on specific potentials affecting the positional dynamics \cite{dauchotDynamicsSelfPropelledParticle2019,guStochasticDynamicsActive2020}, an effective scheme handling the presence of general orientational steering potentials in systems of self-propelled particles is lacking.

Confining and trapping self-propelled particles bear relevance in a wide range of active systems scenarios such as animal foraging for food
\cite{fauchaldUSINGFIRSTPASSAGETIME2003,vilkPhaseTransitionNonMarkovian2022}, chemotaxis of microorganisms and synthetic devices \cite{liebchenPatternFormationChemically2016,starkArtificialChemotaxisSelfPhoretic2018,liebchenSyntheticChemotaxisCollective2018}, as well as optimal control of smart particles \cite{manoOptimalRunandtumbleBased2017,schneiderOptimalSteeringSmart2019,zanovelloOptimalNavigationStrategy2021}.
In one particular scenario a metameric active worm is attracted and steered by the presence of a food resource \cite{duModelMetamericLocomotion2021}, a variant of the general confinement and trapping problem in active matter \cite{aransonConfinementCollectiveEscape2022}.
In this context an important question would be: what is the critical strength of the steering potential,
such that the active particle can find and stay, {\sl i.e.}, be trapped in the vicinity of the food resource.
Mean first exit time (MFPT) is often used to characterize the escape dynamics of confined particles and
the diverging MFPT at a critical potential strength defies simulations, requiring efficient and challenging rare event sampling algorithms \cite{biondini2015introduction}. On the other hand, the infinity of the domain in free space available in the search for food 
poses yet other, no lesser numerical obstacles. Therefore, analytical treatment is essential to address the entrapment dynamics in the regime of critical potential strengths.

Tackling the active Kramers problem analytically relies on effective approximation schemes suitable for specific systems.
Relaxation and fluctuations are two competing ingredients in active Kramers problem and adiabatic elimination of the fast variable to separate relaxation and fluctuation regimes is widely used, {\sl e.g.}, in the local-equilibrium Maxwell-Boltzmann approximation, by comparing translational and rotational time scales \cite{enculescuActiveColloidalSuspensions2011},
as well as in the unified colored-noise approximation \cite{jungDynamicalSystemsUnified1987,marinibettolomarconiStatisticalMechanicalTheory2015} tailored for colored noise 
at small and large correlation time scales. These approximation varieties fall within the purview of the powerful projection operator formalism, applicable in a wide range of contexts \cite{evansMemoryFunctionApproaches1985,gardinerHandbookStochasticMethods2004}.

In this Letter, we investigate the confinement and escape, {\sl i.e.}, the entrapment dynamics of an active particle as exemplified by a metameric worm in steering potential fields \cite{duModelMetamericLocomotion2021}.
By using the projection operator method, we obtain the coarse-grained dynamical equations for general steering potentials in the limit of large orientational fluctuations.
Analytical results for the MFPTs are obtained for radially symmetric potentials allowing us to identify a finite critical steering potential strength for a trapping instability, akin to the counterion condensation phenomenon in the polyelectrolytes theory.
We further verify the analytical results by numerically solving the exact Dynkin equations fulfilled by the MFPTs and show that they effectively coincide with the analytical results.
Our study underpins a quantitative understanding of the behaviors of active particles, subject to steering potentials, by identifying a critical potential strength governing their long-time entrapment behaviors.

\paragraph{Model.}
\label{sec:orgef991b2}
\begin{figure}[htbp]
\centering
\includegraphics[width=.9\linewidth]{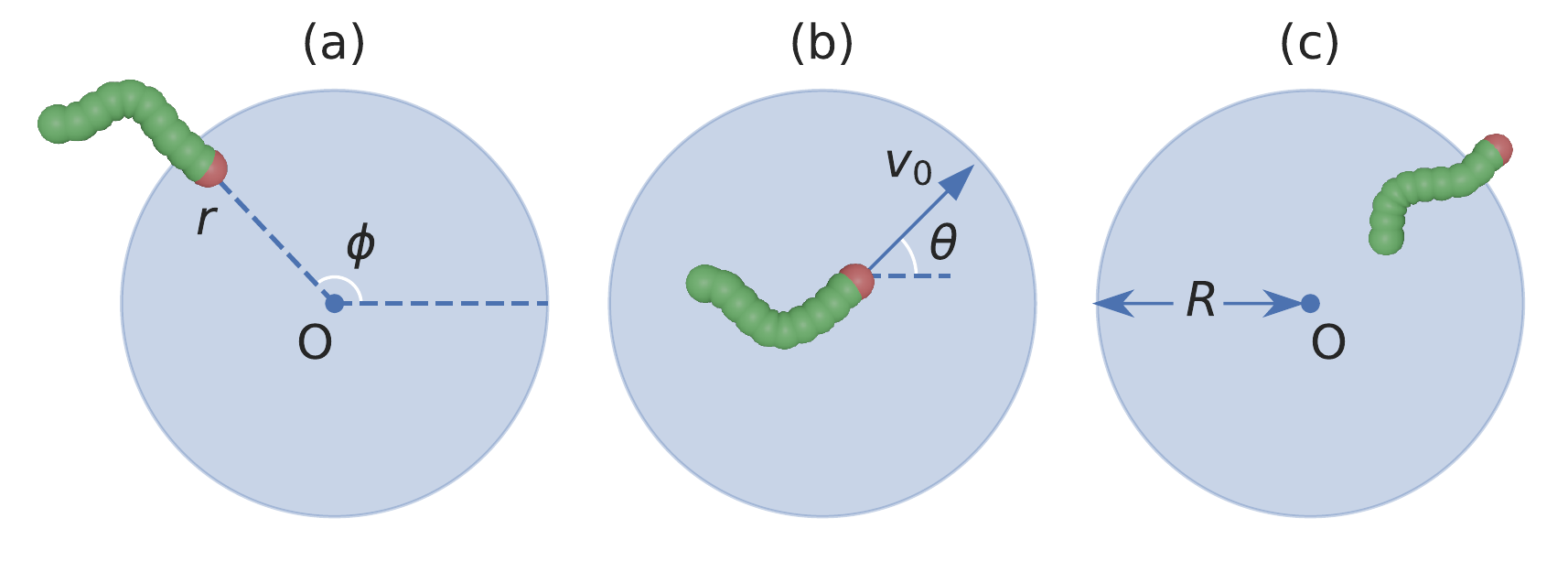}
\caption{\label{fig:schematic}Schematics of the
entrapment [panels (a) and (b)] and escape [panel (c)] of an active worm (an active head and a passive trailing body) in an attractive steering potential field generated by a food resource (blue region). Note that the dimensions of the worm are assumed to be far smaller than the food source radius in text, but have been exaggerated for visual clarity in the figure.}
\end{figure}

To model the confinement of an active particle in a steering potential field, we take a Persistent Turning Walker model \cite{gautraisAnalyzingFishMovement2009,gautraisDecipheringInteractionsMoving2012a} as exemplified by the metameric worm with an active head and a passive trailing body, living in two-dimensional free space \cite{duModelMetamericLocomotion2021}, as illustrated in Fig.~\ref{fig:schematic}. The dynamics of such metameric worm is then given by the dynamics of the active head only. 
The velocity of the active head is of a constant norm \(v_0\) and directed tangentially to its trajectory with orientational angle \(\theta\).
The motion of the worm is steered by a {\sl point source trapping potential} generated by a disk shaped food source of radius \(R\).
We assume the dimensions of the worm to be far smaller than the radius of the food resource and hence the active head will be subsequently deemed to be a point particle.
The position of the active head relative to the center of the food resource is given in polar coordinates \((r, \phi)\), where \(r\) and \(\phi\) are, respectively, the radial distance and the polar angle.
In the Persistent Turning Walker model the curvature dynamics of the worm is governed by a steered Ornstein-Uhlenbeck process \cite{duModelMetamericLocomotion2021}, which is approximated by an overdamped limit curvature dynamics,  corresponding to a large curvature time-decay constant, described by the following Langevin equations
\begin{align}
\frac{\mathrm{d} \mathbf{r}(t)}{\mathrm{d}t} &= v_0 \,\mathbf{t}, \\
\frac{\mathrm{d} \mathbf{t}(t)}{\mathrm{d}t} &= -v_0 \,\mathbf{n}\cdot \bm{\nabla} U(\mathbf{r}; \gamma) \, \mathbf{n} + \sqrt{2D}\,\xi\,\mathbf{n},
\end{align}
where \(\mathbf{r}= r\,(\cos\phi,\, \sin\phi)\), \(\mathbf{t}=(\cos\theta,\, \sin\theta)\) and \(\mathbf{n} = (-\sin\theta,\, \cos\theta)\) are, respectively, the positional, orientational and normal vectors,
\(U(\mathbf{r}; \gamma)\) is the steering potential dependent on position \(\mathbf{r}\), with \(\gamma\) being a parameter representing the potential strength, \(D\) is a constant orientational diffusivity,
and \(\xi\) describes a white noise, \emph{i.e.}, \(\langle \xi(t)\rangle = 0\), \(\langle \xi(t)\, \xi(t')\rangle = \delta(t-t')\).
The term with \(-\mathbf{n}\cdot \bm{\nabla} U\) enables the orientational alignment down the gradient of the steering potential \cite{liebchenPatternFormationChemically2016,starkArtificialChemotaxisSelfPhoretic2018,liebchenSyntheticChemotaxisCollective2018}.

Note that confinement dynamics with similar equations was studied in \cite{aransonConfinementCollectiveEscape2022}, where the focus was placed on the particle trapping and capture
in the limit of vanishing or small fluctuations. While analytical scaling results were given in the deterministic limit, results in the small fluctuations limit were given only numerically. In contrast, here we study the confinement dynamics in the large fluctuations limit and obtain full analytical results for general radially symmetric potentials.

\paragraph{Dynkin equation.}
\label{sec:org85e3e52}
To investigate the entrapment properties, {\sl i.e.}, the  stability of confinement generated by the steering potential, we define the following two criteria: 1) the worm should enter the food resource from outside in a finite time; 2) the worm should stick around infinitely long in the vicinity of the food resource after its entrance.
The confinement is deemed as stable and the worm as trapped only when both of the two criteria are fulfilled.
In what follows MFPT is used to quantitatively characterize the entrapment dynamics and calculating the MFPTs for the worm to go across the boundary of the food resource from the outside [Fig.~\ref{fig:schematic}(a)] and from the inside [Fig.~\ref{fig:schematic}(c)], the two situations are referred to as the {\sl inward} and {\sl outward problems}, respectively.

\begin{figure}[htbp]
\centering
\includegraphics[width=0.8\linewidth]{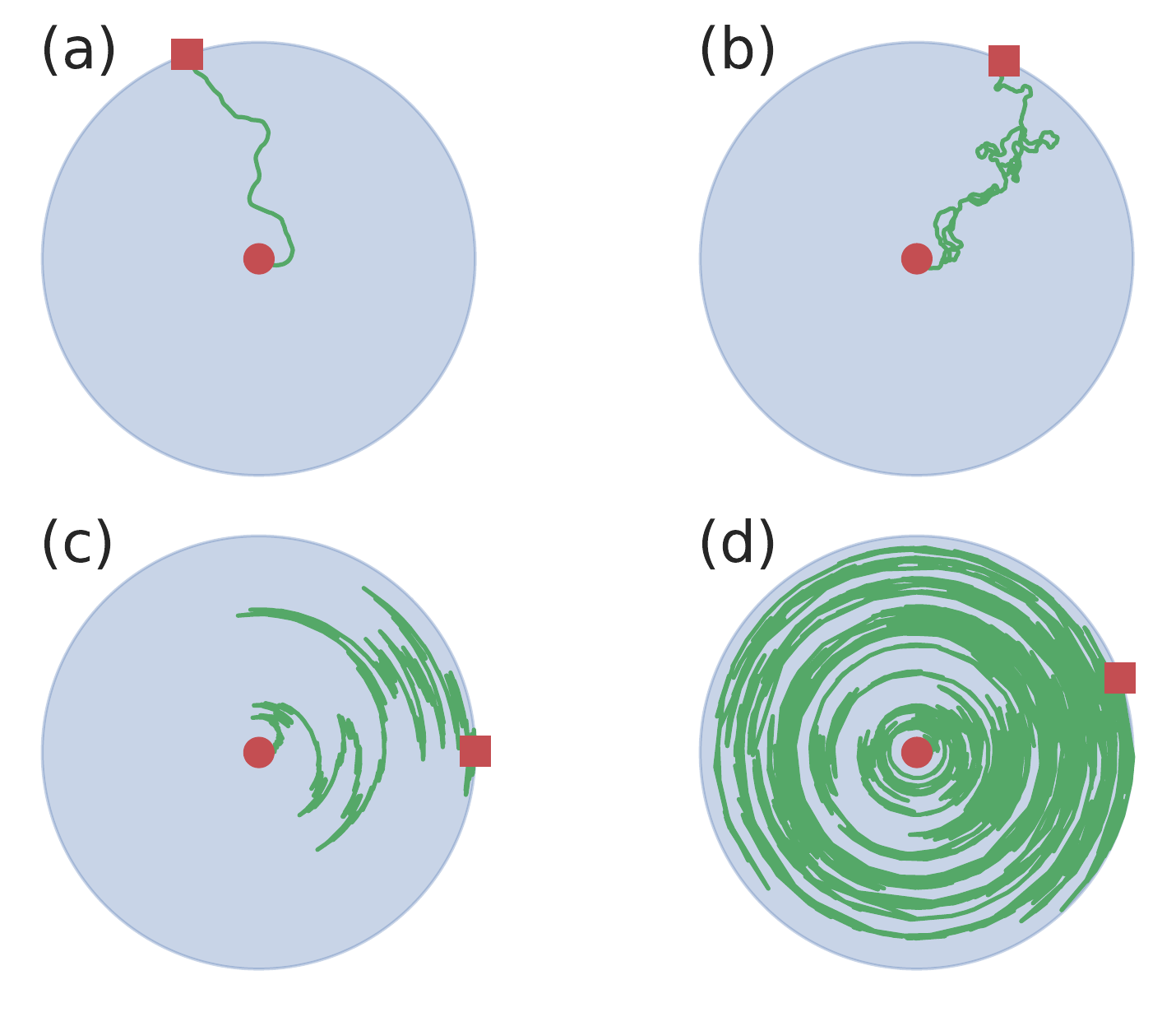}
\caption{\label{fig:trajectory}Typical trajectories in actual positional space coordinates \((r, \phi)\) [panels (a) and (b)] and polar coordinates \((r,\psi)\) [panels (c) and (d)] of an active worm in a steering potential \(U(r) = \gamma \ln r\) (\(\gamma =1\)) with different amplitudes of fluctuations quantified by orientational diffusivity $D$. The worm is initially at the center (circle), with \(\phi=0\) and \(\psi=0\) and finally passes across the boundary (square) of the food resource. In panels (a) and (c), \(D=2\); in panels (b) and (d), \(D=20\).}
\end{figure}

The MFPTs can be obtained directly by solving the corresponding Dynkin equations.
Denote by \(T(r, \phi, \theta)\) the MFPT to cross the boundary of the food resource, which depends on the initial position \((r, \phi)\) and orientation \(\theta\).
By rescaling with the characteristic size \(R\) and time \(R/v_0\), we have the following dimensionless quantities:
\(\tilde{r} = r/R\), \(\tilde{T} = T v_0/R\), \(\tilde{D} = D R/v_0\) and \(\tilde{U}(\tilde{r}, \phi) = U(\tilde{r} R, \phi)\).
For brevity, we will omit the tilde symbol subsequently, as long as this introduces no ambiguity.
Transforming the angular coordinates \((\phi, \theta)\to (\phi', \psi)\) with \(\phi'=\phi\) and \(\psi = \theta-\phi\) then implies \(\partial_{\phi} = \partial_{\phi'} - \partial_{\psi}\) and \(\partial_{\theta} = \partial_{\psi}\).
Denote furthermore \(h(r,\phi,\psi) \equiv \left( \partial_r U - 1/r \right) \sin\psi - [(\partial_{\phi} U - \partial_{\psi} U)/r]\cos\psi\), so that the Dynkin equation can be written in dimensionless form as \cite{suppl}
\begin{align}
\label{eq:dynkin}
\left(\!\cos\psi\, \partial_r + \frac{\sin\psi}{r} \partial_{\phi} + h\, \partial_{\psi} + D \partial_{\psi}^2 \!\right) T(r,\phi,\psi) = -1
\end{align}
with a Dirichlet boundary condition \(T(r,\phi,\psi)|_{r=1}=0\).

\paragraph{Projection operator method.}
\label{sec:org31bd867}

\begin{figure}[tbp]
\centering
\includegraphics[width=0.8\linewidth]{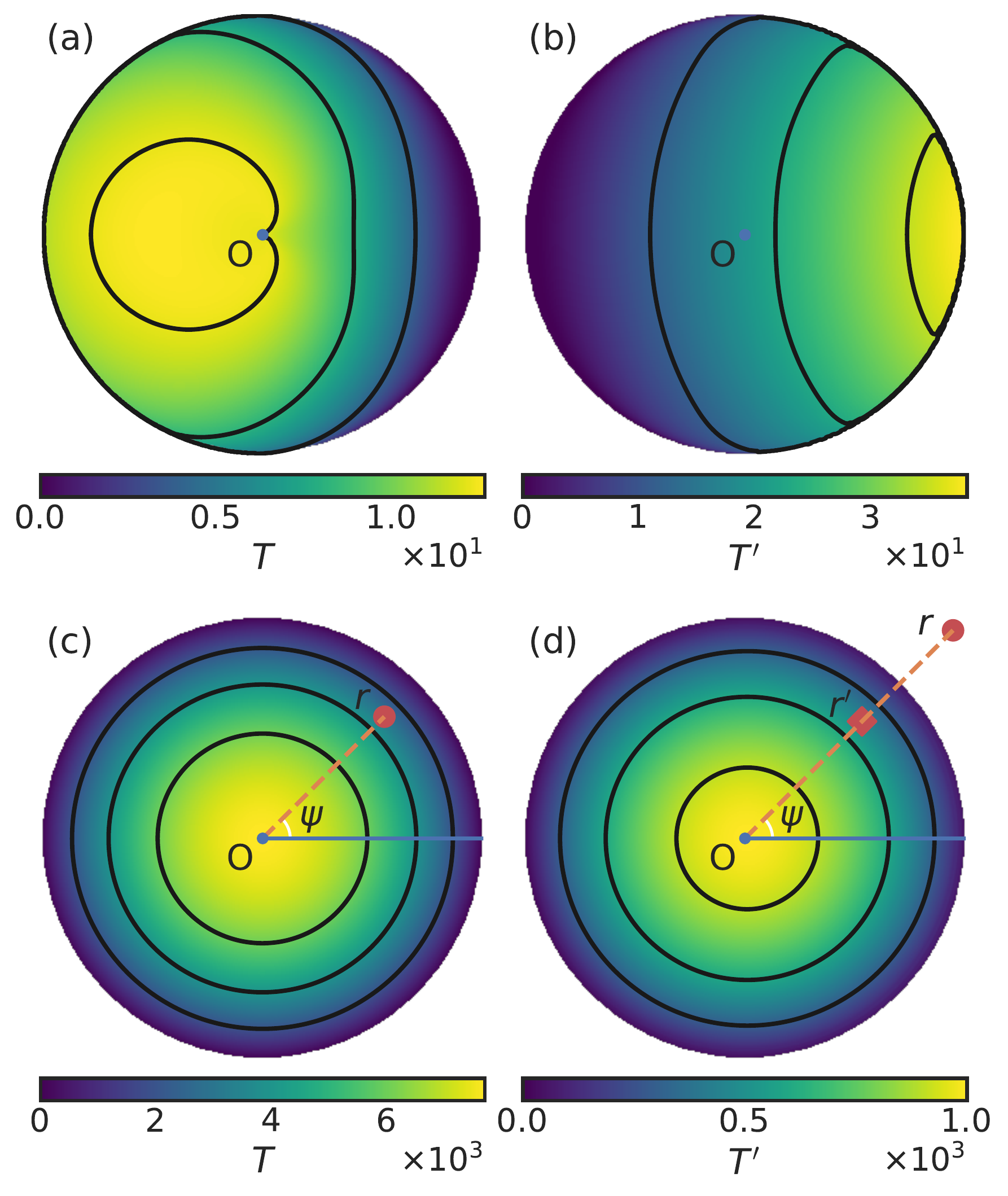}
\caption{\label{fig:symmetric}Density plots of the MFPTs in potential \(U(r)=\gamma \ln r\) with small [\(D=2\) in panels (a) and (b)] and large [\(D=10^3\) in panels (c) and (d)] fluctuations, obtained numerically from the exact Dynkin equation. Panels (a) and (c) show the MFPTs \(T(r,\psi)\) in the outward problem with \(\gamma=1.9\), and panels (b) and (d) show the MFPTs \(T'(r', \psi)\) after the divergence separation \(T'= T/r^2\) and conformal mapping \(r'=1/r\), \(\psi'=\psi\) in the inward problem with \(\gamma=2.1\).}
\end{figure}

To obtain the coarse-grained dynamics, we now apply projection operator method to approximate the Dynkin equation Eq.~\eqref{eq:dynkin}.
We focus on the large fluctuations limit, where analytical results can be obtained.
In this limit, one can expect that the MFPT is independent of the initial orientation.
As shown in Fig.~\ref{fig:trajectory}, the trajectories in polar coordinates \((r,\psi)\) with large orientational fluctuations exhibit a circular pattern, where the effects of the initial orientation are quickly blurred out before the particle can travel a notable distance towards the boundary (\(1/\tilde{D} = v_0/DR \ll 1\) with dimensions  temporarily restored here).
Shown in Fig.~\ref{fig:symmetric} are numerical results of solving the exact Dynkin equation for potential \(U(r) = \gamma \ln r\) with small and large fluctuations.
When fluctuations are small \(D=2\) [Figs.~\ref{fig:symmetric}(a) and \ref{fig:symmetric}(b)], the MFPTs in both outward [Fig.~\ref{fig:symmetric}(a)] and inward [Fig.~\ref{fig:symmetric}(b)] problems clearly depend on the initial value of \(\psi\).
In contrast, when fluctuations are large \(D=10^3\) [Figs.~\ref{fig:symmetric}(c) and \ref{fig:symmetric}(d)], the MFPTs in both outward [Fig.~\ref{fig:symmetric}(c)] and inward [Fig.~\ref{fig:symmetric}(d)] problems are nearly independent of the initial \(\psi\).
Numerical solutions of the inward problem rely on the techniques of conformal mapping and divergence separation, which will be elaborated later.

Numerical examples motivate us to make the operator decomposition \(L = D L_1 + L_2\), where \(L\) is the backward Fokker-Planck operator on the left-hand-side of Eq.~\eqref{eq:dynkin}, with 
\(L_1 \equiv \partial_{\psi}^2\) and \(L_2 \equiv \cos\psi \, \partial_r + 1/r \sin\psi \, \partial_{\phi} + h(r,\phi,\psi) \, \partial_{\psi}\).
Next, define the projection operator \(\mathcal{P}\) by \(\mathcal{P}f(r,\phi,\psi) = 1/(2\pi) \int_0^{2\pi} f(r,\phi, \psi) \, \mathrm{d}\psi, \, \forall f(r, \phi, \psi)\).
By direct calculations, one can show that \(\mathcal{P}\) satisfies 
\(\mathcal{P}^{2} = \mathcal{P}\), \(\mathcal{P}L_1 = L_1 \mathcal{P} = 0\), and \(\mathcal{P} L_2 \mathcal{P} = 0\).
Denote \(\rho(r,\phi) = \mathcal{P}\,T(r,\phi, \psi)\) and \(w(r,\phi, \psi) = (1-\mathcal{P})\, T(r,\phi, \psi)\).
Application of \(\mathcal{P}\) and \(1-\mathcal{P}\) on both sides of Eq.~\eqref{eq:dynkin} then yields
\begin{align}
-1 = \mathcal{P} L_2 \,w, \quad 0 = D L_1 w + (1- \mathcal{P}) L_2\, w + L_2\, \rho.
\end{align}
Subsequent elimination of \(w\) gives a closed equation for \(\rho\),
\(\mathcal{P} L_2 \left[ D L_1 + (1-\mathcal{P}) L_2 \right]^{-1} L_2 \, \rho = 1\), which in the large \(D\) limit implies \(D^{-1} \mathcal{P} L_2 L_1^{-1} L_2 \, \rho = 1\).
Writing this explicitly we remain with
\begin{align}
\label{eq:cg-dynkin-gp}
\partial_r^2 \rho +  \frac{1}{r^2} \partial_{\phi}^2 \, \rho  - A(r,\phi) \, \partial_r \rho + B(r, \phi) \, \partial_{\phi} \rho = -2D,
\end{align}
where
\(A(r,\phi) = 1/\pi \int_0^{2\pi}  h(r,\phi,\psi) \sin\psi \, \mathrm{d} \psi\) and
\(B(r,\phi) = 1/(\pi r) \int_0^{2\pi}  h(r,\phi,\psi) \cos\psi \, \mathrm{d} \psi\).
One can clearly see by rescaling and restoring dimensions that \(\rho \propto D v_0^{-2}\),
which is consistent with previous results for total diffusivity \cite{pototskyActiveBrownianParticles2012}.
Thus by using the projection operator method, we have obtained the coarse-grained Dynkin equation with orientation integrated out in the large fluctuations limit.

\paragraph{Radially symmetric potentials.}
\label{sec:org6bc5959}
To obtain the explicit analytical solutions of the Dynkin equations, we consider radially symmetric steering potentials.
The exact Dynkin equation for a radially symmetric potential \(U(r)\) reads
\begin{align}
\label{eq:dynkin-exact}
\left[ \cos\psi\, \partial_r\! + \!\left(\! \partial_r U \! - \!\frac{1}{r}\!\right) \sin \psi \,\partial_{\psi} \!+\! D \partial_{\psi}^2 \right] T(r, \psi) = \! -1
\end{align}
with a Dirichlet boundary condition \(T(r, \psi)|_{r=1} =0\).
That \(T\) is independent of \(\phi\) can be shown by Fourier transforming the full Dynkin equation or by analyzing directly
the corresponding Langevin equations \cite{suppl}.

According to the projection operation method, see Eq. \eqref{eq:cg-dynkin-gp}, one has 
\(A(r,\phi) = \partial_r U - 1/r\) and \(B(r,\phi) = 0\) for the radially symmetric potential \(U(r)\).
The coarse-grained Dynkin equation is then 
\begin{align}
\label{eq:cg-dynkin-ur}
\partial_r^2 \rho + \left[ \frac{1}{r} - \partial_r U(r) \right] \partial_r \rho = -2D,
\end{align}
with a Dirichlet boundary condition \(\rho(r)|_{r=1} = 0\).

To solve Eq.~\eqref{eq:cg-dynkin-ur}, being of second order, an additional condition is required for both outward and inward problems.
Consider the extended polar coordinates \((-r, \phi)\) which are equivalent to \((r,\phi + \pi)\).
Then \(\rho(-r) = \mathcal{P} T(-r, \psi) = \mathcal{P} T(r, \psi-\pi) = \rho(r)\), where we have used the periodicity \(T(r, \psi + 2\pi) = T(r, \psi)\).
Therefore, we obtain a reflecting boundary condition at \(r=0\), \emph{i.e.}, \(\partial_r \rho(r)|_{r=0} = 0\).
For the inward problem, one can assume a reflecting boundary at a finite radius and subsequently push it to infinity.
Thus we have \(\partial_r \rho(r)|_{r\to\infty}=0\).
The solutions of the coarse-grained Dynkin equations in the outward and inward problems are then, respectively
\begin{align}
\label{eq:solution-general}
\rho^{\text{out},\text{in}}(r) = 2D \!\int_r^1 \!\!\mathrm{d} r_1 \, r_1^{-1} \mathrm{e}^{U(r_1)} \!\!\int_{0,\infty}^{r_1} \!\! \mathrm{d} r_2 \, r_2 \, \mathrm{e}^{ -U(r_2)}.
\end{align}

\paragraph{Trapping instability.}
\label{sec:org554dd4f}
\begin{figure}[tbp]
\centering
\includegraphics[width=\linewidth]{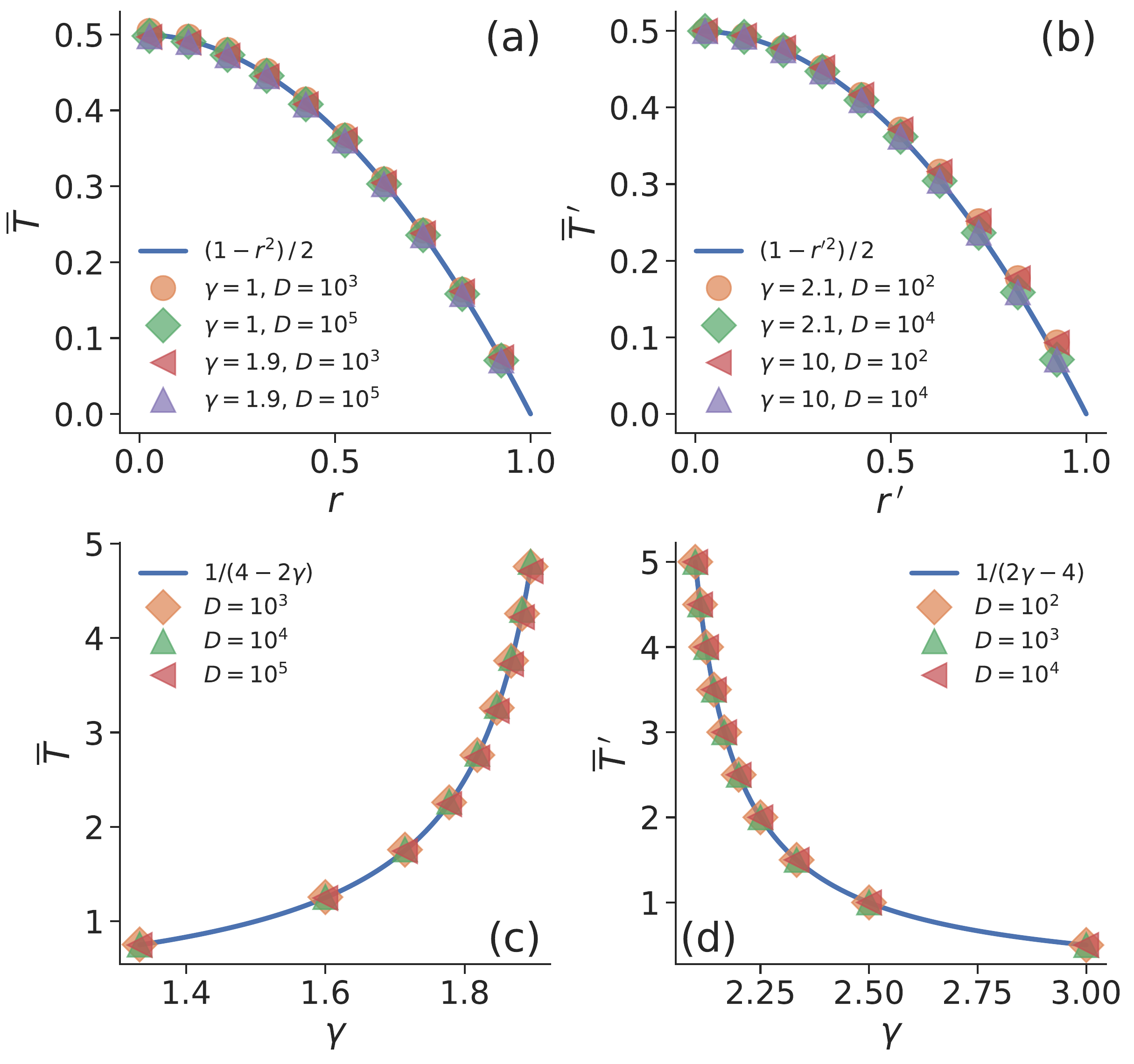}
\caption{\label{fig:n0TrTgamma}MFPTs in potential \(U(r)=\gamma \ln r\) in the outward [panels (a) and (c)] and inward [panels (b) and (d)] problems, obtained analytically (solid lines) from the coarse-grained Dynkin equation and numerically (discrete data) from the exact Dynkin equation. In the inward problem, the MFPTs \(T\) are first transformed by \(T'= T/r^2\) to remove the divergence at infinity, and then - together with the domain of definition - transformed by the conformal mapping \(r\to r'= 1/r\). All the MFPTs are reduced by \(2D\), \emph{i.e.}, \(\overline{T}=T/(2D)\) and \(\overline{T}' = T'/(2D)\). In panels (a) and (b), the numerical MFPTs are along \(\psi=0\). In panels (c) and (d), the numerical MFPTs are at \(r=0\) and \(r'=0\), respectively.}
\end{figure}

To write explicitly the analytical MFPTs in Eq.~\eqref{eq:solution-general}, we consider specifically two kinds of radially symmetric steering potentials: logarithmic and power law.
For a radially symmetric logarithmic potential \(U(r) = \gamma \ln r\),
\begin{align}
\rho^{\text{out},\text{in}}(r) = \frac{D}{2-\gamma} (1-r^2),
\end{align}
where \(\gamma < 2\) in the outward problem and \(\gamma>2\) in the inward problem.
\(\rho^{\text{out}}\) diverges when \(\gamma \geq 2\), and \(\rho^{\text{in}}\) diverges when \(\gamma \leq 2\).
As shown in Figs.~\ref{fig:n0TrTgamma}(c) and \ref{fig:n0TrTgamma}(d) (solid lines), the MFPTs in the outward and inward problems diverge when the potential strength \(\gamma\) approaches \(\gamma=2\) from below and above, respectively.
Combination of the results for the outward and inward problems then yields the critical trapping strength for the whole problem \(\gamma_c = 2\).
When \(\gamma > \gamma_{c}\), the trapping is stable; when \(\gamma < \gamma_c\), the trapping is destabilized.
It is remarkable that the existence of a finite critical \(\gamma_c\) mathematically arising from the factors \(\int_{0,\infty}^{r_1}\mathrm{d}r_2 \, r_2\, \exp[-U(r_2)]\) in Eq.~\eqref{eq:solution-general} resembles the counterion condensation phenomenology in polyelectrolytes theory 
\cite{oosawaPolyelectrolytes1971}.
As shown in in Figs.~\ref{fig:n0TrTgamma}(a) and \ref{fig:n0TrTgamma}(b) (solid lines), the MFPTs, when finite, decrease and increase quadratically with increasing $r$ in the outward and inward problems, respectively.

\begin{figure}[t]
\centering
\includegraphics[width=\linewidth]{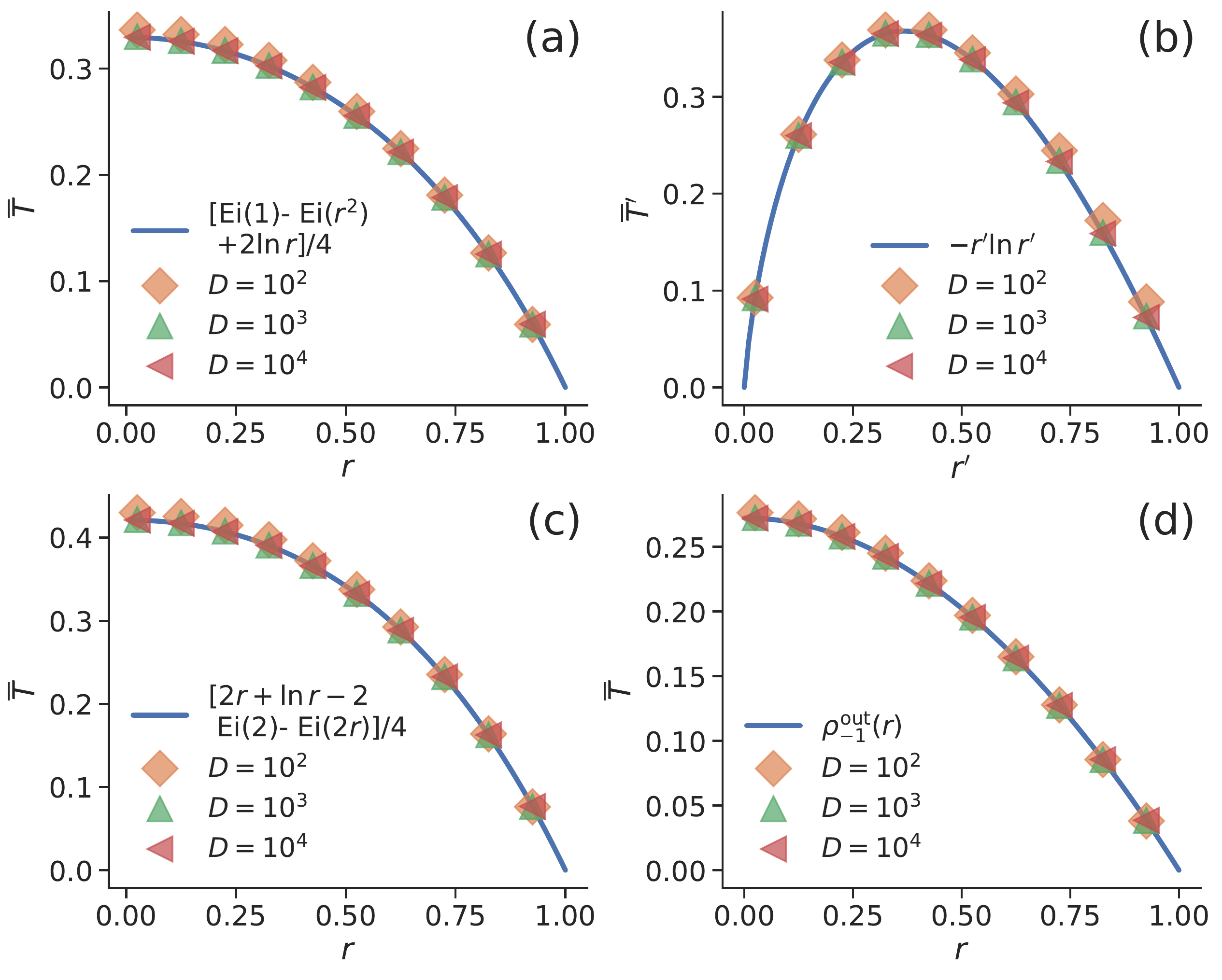}
\caption{\label{fig:n21-1Tr}MFPTs in potentials \(U(r)=\gamma\, r^{\alpha}/\alpha\), \(\alpha=2,1,-1\) in the outward [panels (a), (c) and (d)] and inward [panel (b)] problems obtained analytically (solid lines) from the coarse-grained Dynkin equation and numerically (discrete data) from the exact Dynkin equation. All the MFPTs are reduced by \(2D\), \emph{i.e.}, \(\overline{T}=T/(2D)\) and \(\overline{T}' = T'/(2D)\). The numerical MFPTs are along \(\psi=0\). (a) \(\alpha=2\) and \(\gamma=2\); (b) \(\alpha=2\), \(\gamma=1\), \(T'= T/r^2\) and \(r'= 1/r\); (c) \(\alpha=1\) and \(\gamma=2\); (d) regularized Coulomb-like potential \(U(r) = -\gamma/(r+a)\) with \(\gamma=0.1\) and \(a=0.1\).}
\end{figure}

Moreover, consider radially symmetric power potentials \(U(r) = \gamma \, r^{\alpha}/ \alpha\) with \(\alpha\) being a nonzero real number.
When \(\alpha > 0\), both \(\rho^{\text{out}}\) and \(\rho^{\text{in}}\) are finite, indicating that the worm enters the food resource in a finite time but it wanders away later.
When \(\alpha < 0\), both \(\rho^{\text{out}}\) and \(\rho^{\text{in}}\) are infinite, indicating that the worm cannot enter the food resource in a finite time but will linger infinitely long around the food source if put there.
We write explicitly the analytical results for some values of \(\alpha\), and subsequently corroborate them by explicit numerical calculations. For harmonic potential \(\alpha=2\),
\(\rho^{\text{out}}(r) = D \left[  \text{Ei}\left(\gamma/2\right) - \text{Ei}\left(\gamma \,r^{2}/2\right) + 2 \ln r \right]/\gamma\)
and \(\rho^{\text{in}}(r) = 2D \ln r/\gamma\), where $\text{Ei}$ is exponential integral function.
For linear potential \(\alpha=1\),
\(\rho^{\text{out}}(r) = 2D \left[ \text{Ei}\left( \gamma \right) - \text{Ei}(\gamma \,r) - \gamma + \gamma \, r + \ln r \right]/\gamma^2\) and
\(\rho^{\text{in}}(r) = 2D (-\gamma + \gamma \, r + \ln r)/\gamma^2\).
For Coulomb-like potential \(\alpha=-1\), we use instead a regularized version  \(U(r) = -\gamma/(r+a)\) with \(a\) being a small positive number, since both of the MPFTs in the outward and inward problems in that case diverge.
The MFPT for the regularized Coulomb-like potential in the outward problem obtained from Eq.~\eqref{eq:solution-general} is denoted by \(\rho^{\text{out}}_{-1}\).
For larger $\alpha$, the entrapment is stronger in the inward problem, but weaker in the outward problem [see Figs.~\ref{fig:n21-1Tr}(a) and \ref{fig:n21-1Tr}(c)].

These analytical results for the MFPTs are fully corroborated by numerically solving the corresponding exact Dynkin equations (see the discrete data in Figs.~\ref{fig:n0TrTgamma} and \ref{fig:n21-1Tr}).
However, two problems arise in the calculation of MPFT in the inward problem, {\sl viz.}, 1) the domain is infinite; 2) MFPT diverges at \(r\to \infty\).
A straight-forward way to address these problems would be to manually set a reflecting boundary at a large radius, but this could introduce artifacts in the critical behaviors of MFPT.
Instead, we therefore adopt a method referred to as the {\sl Kelvin transformation}  \cite{nabizadehKelvinTransformationsSimulations2021}.
We first separate the divergence of MPFT by dividing an upper bound of the asymptotic function of MFPT at \(r\to \infty\).
For \(U(r) = \gamma \ln r\), a suitable upper bound is \(r^2\), while for \(U(r) = \gamma r^2/2\), it is \(r\).
Then we use the conformal mapping \(r\to r'=1/r\) to transform the infinite domain into a compact one [see Fig.~\ref{fig:symmetric}(d)] and derive the corresponding exact Dynkin equation. Finally, we numerically solve the exact Dynkin equation after these transformations.
As one can see from Figs.~\ref{fig:n0TrTgamma} and \ref{fig:n21-1Tr}, the reduced MFPTs obtained numerically with different large diffusivities and radially symmetric potentials all collapse onto the lines representing the analytical results, validating the analytical results obtained above.

\paragraph{Summary}
\label{sec:orgd913c0e}
We investigate the entrapment dynamics of an orientationally fluctuating active particle subject to general steering potentials.
By using the projection operator method, we obtain the coarse-grained Dynkin equation with the orientation variable integrated out in the limit of large orientational fluctuations.
For radially symmetric potentials, we explicitly give the analytical solutions of the coarse-grained Dynkin equations.
For logarithmic potential, we find that there exists a trapping instability below a finite critical value of the steering potential strength. It is remarkable that the trapping instability in this context shares the same mathematical origin with the counterion condensation in polyelectrolytes theory. Our findings elucidate the properties of an active orientationally fluctuating Kramers problem by analyzing the  explicit analytical solutions and providing a switch mechanism to regulate the entrapment of steered active particles.

\begin{acknowledgments}
G.D.~and R.P.~acknowledge the funding for the Key Project of the National Natural Science Foundation of China (NSFC) (Grant No.~12034019). F.Y.~acknowledges the support of the National Natural Science Foundation of China (NSFC) (Grant No.~12090054) and the Strategic Priority Research Program of Chinese Academy of Sciences (Grant No.~XDB33030300).
\end{acknowledgments}

%

\ifarXiv
    

\section*{Supplementary material}

## II. DYNKIN EQUATION FOR RADIALLY SYMMETRIC POTENTIALS

For radially symmetric potentials $U(r)$, the Langevin equations are

$$\frac{\mathrm{d}r(t)}{\mathrm{d}t} = \cos\psi, \quad \frac{\mathrm{d}\phi(t)}{\mathrm{d}t} = \frac{\sin\psi}{r}, \tag{S8}$$

$$\frac{\mathrm{d}\psi(t)}{\mathrm{d}t} = \sin\psi\left(\partial_r U - \frac{1}{r}\right) + \sqrt{2D}\,\xi. \tag{S9}$$

Note $\mathrm{d}r/\mathrm{d}t$ and $\mathrm{d}\psi/\mathrm{d}t$ are independent of $\phi$. Hence the MFPT should also be independent of $\phi$, a fact can also be seen from the corresponding Dynkin equation reading

$$\left[\cos\psi\,\partial_r + \frac{\sin\psi}{r}\partial_\phi + \sin\psi\left(\partial_r U - \frac{1}{r}\right)\partial_\psi + D\,\partial_\psi^2\right]T(r,\phi,\psi) = -1. \tag{S10}$$

Separation of variables or Fourier transform with respect to $\phi$ of the above equation shows that $T(r,\phi,\psi)$ is independent of $\phi$. Therefore, the Dynkin equation reduces to

$$\left[\cos\psi\,\partial_r + \sin\psi\left(\partial_r U - \frac{1}{r}\right)\partial_\psi + D\,\partial_\psi^2\right]T(r,\psi) = -1. \tag{S11}$$

## III. NUMERICAL SOLUTION OF DYNKIN EQUATIONS

Here we take $U(r) = \gamma\ln r$ for example to show the preprocessings in numerically solving the Dynkin equations in both outward and inward problems. The numerical solutions are carried out by finite element method, which is most conveniently performed in Cartesian coordinates.

The Dynkin equations for $U(r) = \gamma\ln r$ in the outward and inward problems share the same form

$$\left[\cos\psi\,\partial_r + \frac{\gamma-1}{r}\sin\psi\,\partial_\psi + D\,\partial_\psi^2\right]T(r,\psi) = -1, \tag{S12}$$

except that the defining domains are different. In the outward problem, $r \in [0,1]$; in the inward problem, $r \in [1,\infty)$. The Dynkin equation in the outward problem can be written in Cartesian coordinates ($x = r\cos\psi$, $y = r\sin\psi$) as

$$\left[\left(-\gamma\frac{y^2}{x^2+y^2}+1\right)\partial_x + \gamma\frac{xy}{x^2+y^2}\partial_y\right.$$
$$\left. + D\left(y^2\partial_x^2 + x^2\partial_y^2 - 2xy\partial_x\partial_y - x\partial_x - y\partial_y\right)\right]T(x,y) = -1, \ \sqrt{x^2+y^2} \le 1 \tag{S13}$$

with a Dirichlet boundary condition $T(x,y)|_{\sqrt{x^2+y^2}=1} = 0$.

To numerically solve the MFPT in the inward problem, we have to handle the difficulties of infinite domain and diverging MFPT at $r \to \infty$. Instead of manually setting a reflecting boundary at large radius, we use the so-called Kelvin transformation [2]. To solve the second difficulty, we separate the diverging factor $r^2$ out of $T(r,\psi)$ known from the analytical results, $i.e.$, $T(r,\psi) = r^2\,T'(r,\psi)$. Thus the rescaled MFPT fulfills the equation

$$\left[r^2\cos\psi\,\partial_r + 2\,r\cos\psi + (\gamma-1)\,r\sin\psi\,\partial_\psi + D\,r^2\,\partial_\psi^2\right]T'(r,\psi) = -1. \tag{S14}$$

To solve the first difficulty, we use the conformal mapping $r' = 1/r$ and $\psi' = \psi$, which transforms the infinite domain into a unit disk. We have the Dynkin equation after the conformal mapping

$$\left[-r'^2\cos\psi\partial_{r'} + 2r'\cos\psi + (\gamma-1)r'\sin\psi\partial_\psi + D\,\partial_\psi^2\right]T'(r',\psi) = -r'^2. \tag{S15}$$

Writing it in Cartesian coordinates ($x' = r'\cos\psi'$, $y' = r'\sin\psi'$), we have

$$\left\{\left[-x'^2 - (\gamma-1)y'^2\right]\partial_{x'} + (\gamma-2)x'y'\partial_{y'} + 2x'\right.$$
$$\left. + D\left(y'^2\partial_{x'}^2 + x'^2\partial_{y'}^2 - 2x'y'\partial_{x'}\partial_{y'} - x'\partial_{x'} - y'\partial_{y'}\right)\right\}T'(x',y') = -(x'^2+y'^2), \ \sqrt{x'^2+y'^2}\le 1 \tag{S16}$$

with a Dirichlet boundary condition $T'(x', y')|_{\sqrt{x'^2+y'^2}=1} = 0$.

---


\fi

\end{document}